\documentclass[]{spie}  
\usepackage[]{graphicx}

\title{The infrared imaging spectrograph (IRIS) for TMT: on-instrument wavefront sensors (OIWFS) and NFIRAOS interface} 

\author{David Loop\supit{a}, Vlad Reshetov\supit{a}, Murray Fletcher\supit{a}, Robert Wooff\supit{a}, Jennifer Dunn\supit{a}\\ Anna Moore\supit{b}, Roger Smith\supit{b}, David Hale\supit{b}, Richard Dekany\supit{b}\\ Lianqi Wang\supit{c}, Brent Ellerbroek\supit{c}, Luc Simard\supit{c}, David Crampton\supit{c}
\skiplinehalf
\supit{a}NRC Herzberg Institute of Astrophysics, 5071 W Saanich Rd, Victoria, BC, Canada \\
\supit{b}Caltech Astronomy Department, 1216 E California Blvd, Pasadena, CA, USA \\
\supit{c}TMT Observatory, 2632 E Washington Blvd, Pasadena, CA, USA
}

\authorinfo{Further author information: (David Loop: e-mail: david.loop@nrc-cnrc.gc.ca, telephone: 1 250 363 0022}

\begin{document} 
  \maketitle 

\begin{abstract}
The InfraRed Imaging Spectrograph (IRIS) is a first light client science instrument for the TMT observatory that operates as a client of the NFIRAOS facility multi-conjugate adaptive optics system.  This paper reports on the concept study and baseline concept design of the On-Instrument WaveFront Sensors (OIWFS) and NFIRAOS interface subsystems of the IRIS science instrument, a collaborative effort by NRC-HIA, Caltech, and TMT AO and Instrument teams.  This includes work on system engineering, structural and thermal design, sky coverage modeling, patrol geometry, probe optics and mechanics design, camera design, and controls design.
  
\end{abstract}



\section{INTRODUCTION}
\label{sec:intro}  

We report on the concept study of the on-instrument wavefront sensors (OIWFS) and NFIRAOS Interface of the IRIS science instrument for the TMT observatory, a collaborative effort by NRC-HIA, Caltech, and TMT AO and Instrument teams.

There are a number of papers within this conference describing other facets of the IRIS science instrument; instrument overview (Larkin2010)\cite{Larkin2010}, science case (Barton2010)\cite{Barton2010}, spectrograph design (Moore2010)\cite{Moore2010}, imager design (Suzuki2010)\cite{Suzuki2010}, sensitivities and simulations (Wright2010)\cite{Wright2010}, atmospheric dispersion corrector (Phillips2010)\cite{Phillips2010}, NIR low order wavefront sensor (Hale2010)\cite{Hale2010}

This study has involved patrol geometry, detector alternatives, sky coverage modeling, acquisition and observing scenarios, requirement for high precision astrometry and interfaces with NFIRAOS, the IRIS Science Dewar, and the TMT observatory.  It also included work on probe optics, mechanics, and controls, detector arrays and controllers, instrument rotator and service wrap, and thermal and mechanical design of the OIWFS structure.

The TMT AO team had done prior work in defining the AO system architecture, including low order infrared tip/tilt/focus wavefront sensors for NFIRAOS client instruments.  The top level requirements for the IRIS OIWFS flow down from the overall TMT architecture for adaptive optics and high angular resolution science instrumentation.

The TMT InfraRed Imaging Spectrograph (IRIS) science instrument is located behind the Narrow Field Infrared Adaptive Optics System (NFIRAOS), a facility multi-conjugate adaptive optics system that provides turbulence compensation over a moderately large field of view (2 arcmin) and feeds up to three science instruments.  The reference design of NFIRAOS (figure 1) includes one or more tip/tilt/focus natural guide star wavefront sensors located within each instrument.  The images of natural guide stars will be sharpened by the adaptive optics system to improve the sky coverage for tip/tilt/focus sensing in the infrared with these on-instrument wavefront sensors (OIWFS).

\begin{figure} 
   \centering
   \includegraphics[width=4in]{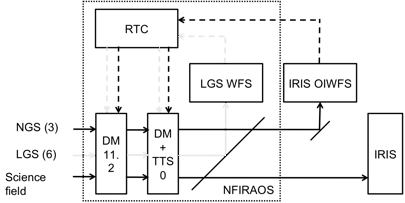} 
   \caption{NFIRAOS reference design}
   \label{fig1}
\end{figure}

The early part of the study involved further development of the requirements to include the NFIRAOS interface and the IRIS science interface, as well as an exploration of possible concept approaches for the OIWFS object selection mechanisms and detectors.  This was a joint effort involving the NFIRAOS team, the IRIS team, and the TMT AO and instrument teams.
  
To obtain acceptable sky coverage and provide good correction of 'blind' modes that are undetectable using the laser guide star wavefront sensors in NFIRAOS, it was determined that the IRIS science instrument requires two tip/tilt (TT) wavefront sensors to perform fast guiding and one tip/tilt/focus (TTF) wavefront sensor to calibrate the focus biases induced by the variations in the range to the sodium layer and laser guide stars.  These OIWFS need to be able to acquire a set of natural guide stars with random positions over the 2 arcmin delivered field of regard (FoR).  This quickly gets complicated if the probes need to reach past and around each other.
  
With the aid of sky coverage modeling by the TMT AO team we were able to identify an eloquent configuration with three identical probes that could each be configured for either TT or TTF operation, arranged symmetrically in a single horizontal plane around the FoR.  Further sky coverage modeling was able to determine that there would be negligible impact to sky coverage if these probes only  patrol half way across the FoR.  Intuitively, this makes sense as a set of natural guide stars arranged around the central science fields is preferred over a set all located to one side.
  
Currently available infrared detectors were explored for read noise versus frame rate performance and we quickly down selected to the Teledyne HxRG as the only currently viable candidate.  The OIWFS detectors are separated from the OIWFS work scope as part of the AO components scope, and it was decided not to explore emerging detector technologies in this phase of study.
  
A Theta-R probe arm configuration was chosen for further development because it is thought to be easier to achieve the desired positioning accuracy and also easier to manage collision avoidance.  The Theta-R probe optics and mechanics were developed using a 2:1 ratio of pixel scales for the TT sensor and the 2x2 subaperture TTF sensor to allow an easy change between TT and TTF operation.
  
The sky coverage modeling tools have been continuously refined through the duration of the study to reflect changes in the low order wavefront sensor AO algorithms, site selection differences, physical optics considerations such as diffraction effects, and the expected performance of the NFIRAOS LGS multi-conjugate adaptive optics.
  
The detector read noise versus frame rate was characterized using sample H2RG detectors, and an innovative co-add multiple readout and windowing schema has been developed to attain very low read noise.  Our modeling analysis indicates that the NFIRAOS sky coverage requirements can be achieved with a lambda-H/2D pixel scale that should also provide sufficient linearity for high precision astrometry.  This provides us with a solid baseline from which we can evaluate future performance and cost trades for emerging infrared detectors.
  
The NFIRAOS interface design was based on the PDR version of the NFIRAOS dual OAP design, which had a well developed interface specification.  A new four OAP design for NFIRAOS is now being developed that may affect our interface design.  The thermal interface design of the interfacing snout shows very good performance, and the mounting sequence has been developed with a set of thermal plugs, gate and diaphragm valves, dry air flushing, and an OIWFS window.
  
Prior art in rotators was explored, and spur gear and friction drive alternatives were analyzed.  A dual motor spur gear drive was chosen.  The rotator is integrated with an external frame structure that extends from the NFIRAOS interface down to the interface with the IRIS science dewar.  A thermal enclosure to maintain the OIWFS volume at -30 C was designed to be nested inside the external frame structure, and an internal platform to mount the OIWFS sensors is supported from flexures close to the delivered focal plane.
  
The service wrap was originally planned to be located close to the NFIRAOS interface, but space limitations forced it to be relocated to the bottom of the science dewar.  A handling cart with lift capability was developed, and the procedure for how IRIS will be offered up to NFIRAOS was also developed.

Operating and observing scenarios for the OIWFS component controller were developed, and high level designs for the hardware and software were generated.
  
\section{INSTRUMENT OVERVIEW}
\label{sec:overview}  

The IRIS science instrument is planned to be mounted on the downward looking port of NFIRAOS.  The maximum allowed mass for IRIS is 5000 kg, and the volume of the IRIS science instrument is limited by the space available between the NFIRAOS frame structure in the vicinity of the downward looking port.  Currently the maximum diameter is 2.48 m and the maximum height  is 5.31 m.  

Each NFIRAOS client instrument must provide its own field derotation, and for the downward looking port this results in IRIS rotating about its optical axis and operating in a gravity invariant environment.

The baseline concept for the IRIS science instrument consists of the following modules as illustrated in the exploded view in Figure 2.
\begin{itemize}
\item Fixed NFIRAOS interface plate and thermal seal, which are used to attach the IRIS science instrument to NFIRAOS
\item IRIS instrument rotator, capable of rotating IRIS to compensate for the natural field rotation arising from the altitude-azimuth telescope mount.
\item IRIS external frame structure, designed to provide support for the OIWFS platform and probes, and the IRIS science dewar. 
\item OIWFS thermal enclosure, designed to keep the OIWFS at a constant working temperature of -30C to limit background emissions.
\item OIWFS platform, located within the thermal enclosure to support the OIWFS wavefront sensors close to the delivered focal plane from NFIRAOS.
\item Three patrolling on-instrument wavefront sensors (OIWFS), which can be individually configured for tip/tilt or tip/tilt/focus operation to deliver natural guide star wavefront pixels to the NFIRAOS real time computer (RTC).
\item Science dewar interface plate, which is used to attach and align the IRIS science dewar to the external frame structure.
\item IRIS science dewar, containing a configurable lenslet / slicer integral field spectrograph, and an infrared imager.
\item Service wrap subassembly, capable of supplying power, communications, and cooling services to the rotating IRIS science instrument from the fixed Nasmyth platform.
\end{itemize}

\begin{figure} 
   \centering
   \includegraphics[height=8.5in]{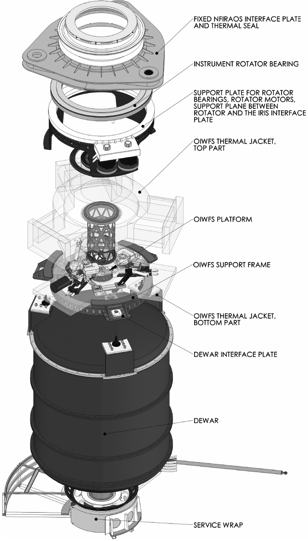} 
   \caption{IRIS science instrument overview}
   \label{fig2}
\end{figure}

\section{SKY COVERAGE}
\label{sec:sky}

For NFIRAOS, the tip/tilt sky coverage describes the possibility of locating natural guide stars that enable tip/tilt compensation to meet or exceed the overall wavefront error performance requirement.

Both a Zernike-based geometric optics code and a higher-fidelity physical optics model have been used in the sky coverage simulations.  The geometric simulator was developed first (Clare2006)\cite{Clare2006}.  The OIWFS design choices developed using this tool include; the patrol arm restrictions, the number and type of OIWFS, and the spectral bandpass.  

The geometric optics model does not simulate the full details of the NGS image on the OIWFS focal plane, and cannot accurately estimate the impact of measurement noise, higher-order wavefront aberrations, or detector pixel size upon OIWFS performance.  The physical optics model was developed to address these issues (Wang2009)\cite{Wang2009}.  The design choices derived using this more detailed approach include improvements to the pixel processing and control algorithms, re-optimizing the DM fitting FoV to better sharpen the NGS images, and optimizing detector pixel sizes to balance the effects of measurement noise and non-linearity.

\subsection{Patrol Arm Restrictions}
In the current baseline design, the OIWFS use Theta-R probe arms with pivots outside of the patrol field.  These arms are on the same horizontal plane, and therefore they cannot cross over each other.  In addition, these arms are not allowed to enter the science field to prevent vignetting.  

It was suggested that each patrol arm could be able to switch between TT and TTF mode and only patrol part way across the FoR.  We modeled these ideas and found that with functional change capability limiting the patrol arm range to about half way across the patrol FoR does not measurably penalize the sky coverage.

\subsection{Number and Type of OIWFS}
With the low effective read out noise of the detectors considered, it was of interest to study the sky coverage obtained with 3 TTF sensors to reduce the complexity of the opto-mechanical design (in comparison with having switchable TT/TTF functionality).  However, simulations have shown that using 3 TTF WFS degrades sky coverage significantly. 

Figure 3 shows the sky coverage for various numbers of TT and TTF WFSs for a given detector.  It can be seen that using 1 TTF and 2 TT gives the best performance for a wide range of sky coverage.  The switchable TT/TTF functionality allows this choice to be revisited for specific observations.

\begin{figure} 
   \centering
   \includegraphics[width=5in]{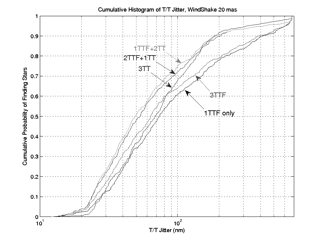} 
   \caption{Number and type of OIWFS}
   \label{fig3}
\end{figure}

\subsection{Wavefront Sensing Passband}
Employing both J+H bands for wavefront sensing is possible with the detectors available, and can help improve the sky coverage substantially without significantly complicating the optical design.  The Teledyne H2RG HgCdTe detector has similar quantum efficiency in J and H bands, and yet the partially corrected PSF provided by the NFIRAOS higher-order wavefront correction has a significantly better Strehl in H band.

A significant improvement is achieved using J+H due to 1) improved signal to noise ratio due to more light collected and 2) the better Strehl ratio in H band.  We have chosen J+H band wavefront sensing as the baseline.

\subsection{Off Zenith Sky Coverage} 
Modeling the sky coverage for off-zenith telescope pointing shows the RMS wavefront error increases significantly for zenith angle greater than about 30 degrees.  The reasons for this are twofold:  On the one hand, the high order correction on the NGS WFS images degrades with increasing zenith angle, which reduces the OIWFS SNR.  On the other hand, the increased off-zenith turbulence and reduced isoplanatic angle also causes the correction to be less effective.  The performance degradation is fairly mild at zenith angles below 30 degree, but becomes very significant at zenith angles above 45 degree.

\subsection{Algorithm Upgrades} 
With the new physical optics sky coverage code we made several improvements to the processing algorithms and control parameters.

We found that the error in plate scale modes can be reduced by employing a type II controller, which is composed of a double integrator and a lead filter to provide phase margin.  The improved error rejection of the type II controller at lower temporal frequency helps to reduce the residual error in the plate scale modes, which are of relatively low temporal frequency.  The phase lead, frequency, and overall gain of the double integrator are optimized during the post processing step for each sampling frequency for each NGS asterism to ensure a 45 degree phase margin, and to balance the error rejection and noise propagation.  The overall gain at high sampling frequencies is significantly reduced to suppress the error due to noise.
	
The NFIRAOS tip/tilt stage prototype shows an improved bandwidth of 90 Hz compared to the requirement of 20 Hz that was used to tune the original woofer/tweeter control.  With this faster woofer, the original woofer/tweeter control algorithm loses phase margin and is hard to make stable.  We modified it to use the same type II controller as the plate scale control loop, which is composed of a double integrator and a lead filter.  The error rejection is significantly improved for low temporal frequency tip/tilt errors for this new approach. The residual error at low sampling frequencies is significantly reduced.  Similar improvements are also obtained for the tip-tilt errors due to atmospheric turbulence.
	
\subsection{DM Fitting FoV} 
In the higher-order LGS control loop, the deformable mirror actuator commands are selected to optimize the residual RMS wavefront error within a specified ÒDM fittingÓ FoV.  We found that the sharpening of the NGS guide stars can be improved by optimizing the DM fitting field for a FoV of 30 arcsec in diameter, slightly larger than required for the IRIS science FoV.  This improves the sky coverage significantly by improving the magnitude limit of the OIWFS, but it does not penalize the Strehl obtained on the science FoV appreciably.

\section{NFIRAOS INTERFACE}
We have developed concept designs for the macro-mechanics in conjunction with the IRIS design team including,
\begin{itemize}
\item the fixed NFIRAOS kinematic interface plate and thermal seal
\item instrument rotator bearing, gear ring and motors(2)
\item the support frame between the rotator and the IRIS interface plate
\item the OIWFS thermal structure
\item the OIWFS support frame and probe platform
\item the IRIS kinematic interface plate
\item the service wrap
\end{itemize}

\subsection{NFIRAOS Interface Plate} 
The NFIRAOS interface mounting plate shown in figure 4 provides a kinematic mount to the mounting pads provided on NFIRAOS.  As IRIS is lifted and comes in close proximity to the NFIRAOS mounting pads, the OIWFS assembly will be guided into a repeatable mounting position by two tapered dowel pins against hole and slot features to satisfy the mounting repeatability requirement.
\begin{figure} 
   \centering
   \includegraphics[width=5in]{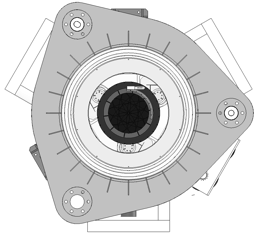} 
   \caption{NFIRAOS interface plate, top view}
   \label{fig4}
\end{figure}

\subsection{NFIRAOS Thermal Seal} 
The thermal interface between IRIS and NFIRAOS serves several functions.  In the observation mode it provides a thermal seal that eliminates infiltration and thermal plumes into the NFIRAOS enclosure, and it must accommodate the rotation of IRIS with respect to NFIRAOS.  During the installation of IRIS on NFIRAOS it must protect the IRIS-OIWFS window from condensation.

The design of the IRIS-NFIRAOS thermal interface was driven by the necessity to have a static seal between IRIS and NFIRAOS.  A dynamic rotary seal between the two mating ports would have made the requirements for alignment and installation of IRIS much more stringent, and increased the risk of seal damage.

The seal starts with a cold labyrinth formed by two aluminum weldments.  The top labyrinth plate is stationary and is pressed by compression springs against the cold plate of NFIRAOS.  The bottom labyrinth plate is fixed to the OIWFS buried cold plate.   The cold labyrinth receives the heat flow through two sets of seals;  the first path is through two rotary lip seals, and the second one is though the static face seals.

First approximation thermal finite element analysis was performed to calculate the heat influx though this seal arrangement.  The analysis indicated a negligible amount of the heat flow.

If IRIS were to be attached to NFIRAOS without cooling it down to -30 ¡C, it would create thermal perturbations that would take NFIRAOS out of operation for several hours until the OIWFS cavity is cooled to -30 ¡C.  To shorten the time between the attachment of IRIS and the start of operation, a diaphragm mechanism has been added to the IRIS thermal interface design that allows for cold attachment.  It consists of a custom-made iris type diaphragm and a window underneath it.  The volume between the diaphragm and the window will be flushed with dry air and to prevent condensation.  

\subsection{Instrument Rotator} 
The IRIS rotator mechanism consists of the NFIRAOS interface plate, rotator bearing, gear ring bolted to the interface plate and a pair of AC motors.  To avoid gear backlash, the motors are exerting torque in opposite directions, thus one of the motors will always be maintaining a preloading back-driving force against the forward-moving motor.  The advantages of the direct drive AC motors include very low torque ripple, high torque densities, and increased reliability over brush-type DC motors.  This rotator design is very similar in all principal elements to the LBT rotator (Ashby2008)\cite{Ashby2008}.  The LBT rotator design was adopted because of similar requirements in terms of geometry, supported weight and repeatability.  The components to be used in the IRIS rotator (motors, gear, rotator bearing) were selected based on the IRIS specifications, and are different from those used for the LBT rotator.

Friction drive was considered as an alternative to the spur gear drive.  Generally, friction drives are widely used in telescope and instrument rotators (Leger2003)\cite{Leger2003}, (Leger2004)\cite{Leger2004}, (Wang2004)\cite{Wang2004}, (Hammerschlag2006)\cite{Hammerschlag2006}, (Yang2008)\cite{Yang2008}.  Among the advantages of friction drives is the absence of backlash (therefore only one motor is needed), the absence of any significant short-term irregularities, and relatively low cost.  The disadvantages of friction drives include a necessity for high quality of surface disks, high contact pressures requiring high uniformity surface heat treatment, sensitivity to contamination, such as insect infestations, driving rollers requiring very rigid and precise bearing support.  The initial assessment shows that a spur gear drive will be lighter, more compact, and will have fever critical custom-made parts than a friction drive.  The relative simplicity of the spur gear drive will translate into reduced schedule and performance risks.  The rotator spur gear drive system is shown in figure 5.
\begin{figure} 
   \centering
   \includegraphics[width=5in]{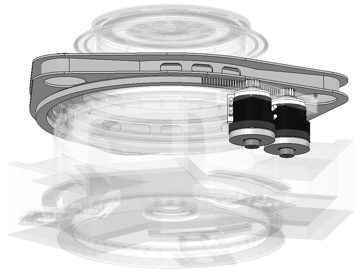} 
   \caption{Spur gear rotator drive system}
   \label{fig5}
\end{figure}

\subsection{OIWFS Thermal Enclosure}
To reduce thermal background, the OIWFS internal enclosure will be actively cooled prior to operation to -30¡C using air handling units, and kept at that temperature during operation using cold-wall insulation.  The OIWFS cold-wall insulation concept is similar to the NFIRAOS cold-wall design.  It is a multi layer configuration with an embedded actively cooled metal sheet within two thermally insulating layers clad with a fiberglass composite.
  
The thermal insulation jacket is separated into five components: bottom plate, top jacket and three access plates.  The bottom plate and the top jacket are supported on the external structural components. The access panels are removable, and supported on the top jacket.

\subsection{OIWFS Probe Platform}
The OIWFS platform (figure 6) supports three probe arms, the stationary optical components of the probe arms, the OIWFS cameras, and the OIWFS window.  Probe arms will be integrated and tested on the OIWFS platform, outside of the instrument.

The platform is attached to the dewar interface frame with titanium flexures. There are six struts in total; on both of their ends each strut has a flexure equivalent of a ball-socket joint.  Altogether they form a hexapod: a kinematically constrained structure.  The hexapod mount insulates the OIWFS platform from thermal distortions of the outside structure.
\begin{figure} 
   \centering
   \includegraphics[width=5in]{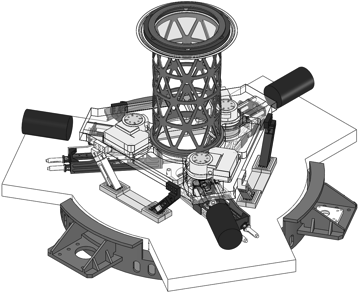} 
   \caption{OIWFS probe platform}
   \label{fig6}
\end{figure}

\subsection{Science Dewar Interface}
The IRIS Science Dewar is mounted to the OIWFS assembly using a design identical in concept to the NFIRAOS-IRIS mounting interface.  The mechanical interface consists of three pads on the OIWFS-Science Dewar interface plate that have been machined flat and drilled with clearance holes for the bolts.  The Science Dewar is indexed relative to OIWFS using two tapered dowel pins located on the dewar mounting pads which engage with ÒholeÓ and ÒslotÓ features on the OIWFS structure.  These features are independently adjustable (via jack screws) to align the OIWFS and Science Dewar optical axes.

\subsection{Service Wrap}
The requirements call for a 270 degree range of rotation for the science instrument, and a fairly traditional rotating chain track cage arrangement has been developed.  The full range of services required by the IRIS science instrument have not yet been developed, and a set of power, electronics, air, coolant, and gaseous helium services have been assumed.

\section{ON-INSTRUMENT WAVEFRONT SENSORS}
The first part of the concept study explored architectural alternatives for the natural guide star wavefront sensors, and the following object select mechanisms were considered.
\begin{itemize}
\item Robot placed pickoff and path length mirrors
\item Tip/tilt mirror tiling of the patrol focal plane
\item Theta-Phi probes, with 2 rotation stages
\item Theta-R probes, with 1 rotation + 1 linear stage
\end{itemize}
The requirements for positioning accuracy, dithering, and non-sidereal tracking pushed the concept choice towards movable probes, and the Theta-R probe concept was chosen over the Theta-Phi probe concept because of accuracy and collision concerns.  The probe pickoffs are located at the delivered focal plane approximately 750 mm from the NFIRAOS mechanical interface with the Theta-R probes patrolling the 2 arcmin curved focal plane field of regard (FoR).  This arrangement is acceptable to the IRIS science dewar design team, with their first optic located approximately 100 mm after the delivered focal plane.

The plate scale at the delivered focal plane is 2.18 mm/arcsec, which translates the 2 mas RMS positioning accuracy requirement into +/-4.36 um 1 sigma radius accuracy.  With arbitrary location of the tip/tilt (TT) and tip/tilt/focus (TTF) natural guide stars it initially looked like we would need to have separate horizontal patrol planes so the TT and TTF probes could reach past each other.  A breakthrough came with the realization that we could configure each probe for either TT or TTF function by inserting either an imager lens or a 2x2 lenslet.  The lens and lenslets differ in focal length by a factor of 2 to give the appropriate scales at the detector.  A trombone mirror arrangement is used to put the NFIRAOS exit pupil on the lens or the lenslets as well as compensate for the change in path length around the patrol field.  This allowed us to have three identical Theta-R probes at 120 degree spacing around the 2 arcmin field of regard.

When this concept was analyzed for sky coverage a further breakthrough came with the realization that each probe only needs to reach half way across the 2 arcmin field of regard to get a suitable natural guide star asterism around the on-axis science field.  Figure 7 shows a plan view of the patrol geometry for this arrangement.
 \begin{figure} 
   \centering
   \includegraphics[width=5in]{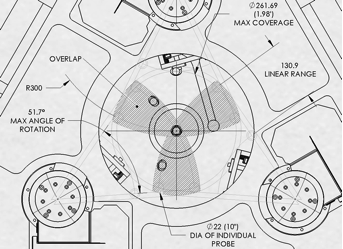} 
   \caption{OIWFS patrol geometry}
   \label{fig7}
\end{figure}

\subsection{OIWFS Probe Optics}
The Theta-R probe optics design was developed to reflect the changeable TT and TTF functionality, as illustrated in figure 8.  

The probe pickoff tip incorporates a converter lens, pickoff fold mirror and circular field stop sized to fit within the detector pixel area.  The collimator lens on the probe arm incorporates focus motion to compensate for defocus of the curved NFIRAOS delivered focal surface and maintain probe image quality.  

The trombone mirrors at the back of the linear stage are adjustable to maintain the pupil patch focus for TT and TTF configurations.  The trombone mirrors move at half the rate of the probe tip to maintain path length, but the TT and TTF function change necessitates separate linear actuators for the probe arm and trombone mirrors rather than a 2 to 1 screw pitch arrangement.  

A pair of fold mirrors translates the optical path through the rotation stage, with the remainder of the path and detector held stationary.  Each probe includes a separate atmospheric dispersion corrector (ADC) to compensate for telescope elevation change.  The TT imager mono lens and the TTF 2x2 lenslet array are inserted or removed as required.
\begin{figure} 
   \centering
   \includegraphics[width=5in]{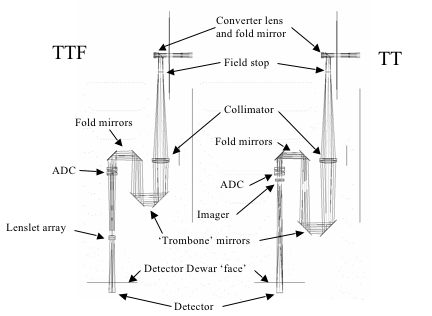} 
   \caption{OIWFS optics design}
   \label{fig8}
\end{figure}

\subsection{OIWFS Probe Mechanics}
The mechanics of the probe arm consist of the following major subassemblies: linear stage, collimator actuator, rotation stage, ADC assembly, OIWFS camera assembly.  This modular structure is advantageous because it allows assembly and testing of these major subcomponents independently of each other.

The linear stage is a custom design that incorporates the required features.  The probe tip carriage and trombone mirrors carriage are moved in a common housing on separate linear guides and lead screws.  Each carriage is carried by four blocks (two on each side), and the blocks are preloaded.  Four blocks per carriage are used to average out the non-repeatable variations and decrease parasitic angular motions along the length of travel.  It is expected that deviations from straightness will be measured and calibrated out in the assembly and integration process.  The guides are bolted to the housing steel plates to match the thermal expansion coefficient.  The linear guides are located in the same horizontal plane with the probe arm optical axis. This location minimizes Abbe errors in the vertical plane, and having two linear guides and four blocks per carriage reduces Abbe errors in the horizontal plane.

The position of the probe arm is measured with a high resolution magnetic absolute encoder mounted on the probe arm carriage.  The encoder head is stationary on the probe arm housing, and the scale moves with the carriage.  A sectional view of the linear stage is shown in figure 9.
\begin{figure} 
   \centering
   \includegraphics[width=5in]{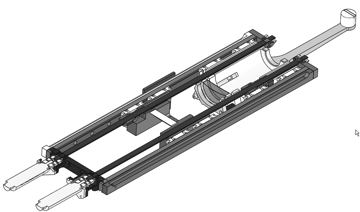} 
   \caption{Sectional view of OIWFS linear stage}
   \label{fig9}
\end{figure}

The collimator lens needs to be translated to compensate for delivered focal plane curvature defocus with high repeatability.  These motion requirements are not difficult to achieve, but the space envelope, however, does present a challenge.  The collimator lens is translated on two linear miniature stroke guides with two DC motors used as actuators.  It is necessary to use two motors in order to balance out the parasitic moments, and to avoid tilting the collimator element with respect to the optical axis.

The probe arm rotator utilizes an off-the-shelf rotation stage.  It provides high-precision angular positioning accuracy combined with high load capacity.  This stage is constructed from stainless steel, with rotation accuracy ensured by ground bearing surfaces.  A double row of preloaded bearings allows for high off-center loads, while typical eccentricity of rotation is only 1.4 um.  This stage is driven by a precision ground and hardened worm gear with self-compensating preload and is equipped with a direct reading optical encoder attached to the moving platen.

The rotation stage is mounted directly to the OIWFS platform.  The thermal expansion mismatch between the rotation stage and the OIWFS platform is mitigated with flexure reliefs machined into the platform.  The linear stage is connected to the bottom side of the rotation stage by a hollow rotator shaft that houses a rotating fold mirror.  The ADC assembly protrudes into the rotator shaft with the stationary fold mirror mount.  The probe camera and ADC assembly are mounted in their fixed orientations to the OIWFS platform.

The ADC assembly is a stationary module.  It can be put together and aligned outside of the OIWFS platform.  Within the OIWFS platform it will be aligned to the optical path as a module.  The ADC assembly contains the stationary fold mirror, two ADC elements, the imager lens, and the lenslet array (figure 10).
\begin{figure} 
   \centering
   \includegraphics[width=3.5in]{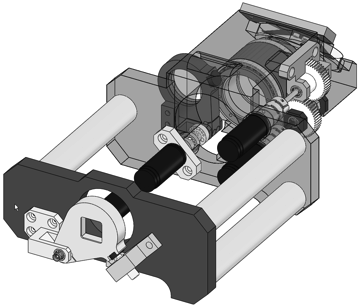} 
   \caption{ADC assembly}
   \label{fig10}
\end{figure}

The first element in the optical path is the stationary fold mirror.  It is held in place by a flat spring pressing it against three pads.  The pads are machined into the ADC housing. Sideways movement of the mirror is limited within the clearance between the mirror and the housing bore.  Each of the two ADC elements can be rotated independently.  Each ADC optical cell is mounted into a four point contact self-preloaded bearing, and two DC motors drive their respective ADC optical cells through a spur gear drive.

The OIWFS probe camera consist of a windowed dewar, cooled to 77 K, that houses the baseline H2RG infrared detector and J+H passband filter (figure 11).  The detector team at Caltech developed a innovative multiple co-add readout schema for the H2RG to achieve very low read noise, and a detector controller to send pixel photon counts to the NFIRAOS real time controller (RTC) (Hale2010) \cite{Hale2010}.
\begin{figure} 
   \centering
   \includegraphics[width=4.5in]{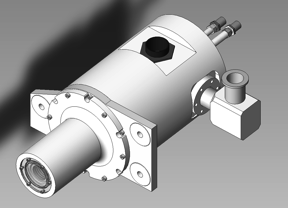} 
   \caption{OIWFS probe camera}
   \label{fig11}
\end{figure}

\subsection{OIWFS Component Controller}
The IRIS OIWFS component controller will provide the following functions:
\begin{itemize}
\item Control the rotator mechanism for IRIS.
\item  Mechanical pointing control of each WFS relative to the delivered output NFIRAOS focal plane.  This allows higher-level systems to perform acquisition of IR natural guide stars (NGSs), dithering of IR NGSs (in coordination with NFIRAOS to make a predictable pointing offset of the science image relative to the IRIS spectrograph and imager), and tracking of non-sidereal objects (to maintain the science image at the same location relative to the IRIS spectrograph and imager).
\item Atmospheric dispersion correction, to maintain both the relative pointing of sensors viewing different colour NGS and the potentially different again colour science target.
\item Temperature control of the IRIS OIWFS.
\end{itemize}
\begin{figure} 
   \centering
   \includegraphics[width=6in]{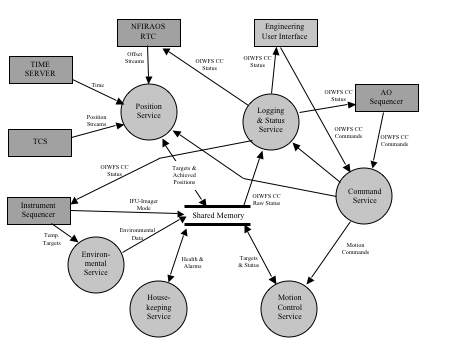} 
   \caption{OIWFS CC software decomposition}
   \label{fig12}
\end{figure}

The program and data structures and data flow for the OIWFS component controller are shown in figure 12.  The data flow descriptions are detailed below.\skiplinehalf
OIWFS Position Streams (from TCS)
\begin{itemize}
\item OIWFS Probe position target streams from TCS (1-3 probes, X, Y and time)
\item OIWFS ADC target streams from TCS (1-3 ADCs, orientation, power and time)
\item IRIS Rotator position target stream from TCS (angle and time)
\end{itemize}
Commands and Status
\begin{itemize}
\item AO Sequencer sends the usual commands: initialize, index, move, park, configure, follow
\item CC responds with command acknowledgement and ongoing status
\item CC provides indication of health to AO Sequencer
\item CC reports status of tracking loops (i.e. tracking or not Tracking plus current position errors)
\end{itemize}
Temperature Control and Status (from/to IRIS Instrument Sequencer)
\begin{itemize}
\item Temperature set points for OIWFS chamber and OIWFS Detectors
\item Achieved temperatures for a selection of OIWFS components
\end{itemize}
IRIS Operating Mode (from IRIS Instrument Sequencer)
\begin{itemize}
\item IFU or Imaging mode (for mechanism pointing models)
\end{itemize}
Engineering Data (to Data Management System)
\begin{itemize}
\item TBD
\end{itemize}
OIWFS Offsets (from NFIRAOS RTC)
\begin{itemize}
\item IRIS Rotator offsets (angle)
\item OIWFS ADC corrections (?, a/b)
\item OIWFS Probe position offset targets (maybe)
\end{itemize}
OIWFS Status (to NFIRAOS RTC)
\begin{itemize}
\item Achieved OIWFS Detector temperatures
\item Achieved IRIS Rotator position
\end{itemize}
OIWFS Engineering Commands and Status (from/to Engineering User Interface)
\begin{itemize}
\item Target OIWFS commands, modes, positions, temperatures, etc. to CC 
\item Command status and achieved modes, positions, temperatures, etc.
\end{itemize}

The TME AO and instrumentation teams defined the basic observation workflow and the design team extended these to include all the required operating and observing scenarios; cool down, cold start, warm start, acquisition, tracking a sidereal object, tracking a non-sidereal object, nodding, dithering, ending observation, park, shutdown, warm up, and calibrations.

\acknowledgments     
 
The authors gratefully acknowledge the support of the TMT partner institutions. They are the Association of Canadian Universities for Research in Astronomy (ACURA), the California Institute of Technology, and the University of California. This work was supported as well by the Gordon and Betty Moore Foundation, the National Research Council of Canada, the Association of Universities for Research in Astronomy (AURA) and the U.S. National Science Foundation. 


\end{document}